\documentclass[tightenlines,twocolumn,nofootinbib,prd,showpacs,aps,epsf,amsfonts,amssymb,amsbsy]{revtex4}

\usepackage{graphics,bm}
\usepackage{epsfig}
\usepackage{graphicx}

\newcommand{\beq}{\begin{equation}}
\newcommand{\eeq}{\end{equation}}
\newcommand{\bqa}{\begin{eqnarray}}
\newcommand{\eqa}{\end{eqnarray}}

\def\square{\vcenter{\vbox{\hrule height.4pt
          \hbox{\vrule width.4pt height4pt
          \kern4pt\vrule width.3pt}\hrule height.4pt}}}

\begin{document}


\setlength{\unitlength}{1mm}
\author{Jens O. Andersen} 
\affiliation{Department of Physics, 
Norwegian University of Science and Technology, N-7491 Trondheim, Norway} 
\affiliation{
Niels Bohr International Academy, Niels Bohr Institute and Discovery Center,
Blegdamsvej 17, DK-2100 Copenhagen, Denmark}
\title{Thermal pions in a magnetic background}

\date{\today}

\begin{abstract}
We use chiral perturbation theory for $SU(2)$ to
compute the leading loop corrections to the thermal mass of the pions 
and the pion decay constant in 
the presence of a constant magnetic field $B$.
The magnetic field gives rise to a splitting between 
$M_{\pi^0}$ and $M_{\pi^{\pm}}$ as well as $F_{\pi^0}$ and $F_{\pi^{\pm}}$.
We also calculate the 
free energy and the quark condensate to next-to-leading order
in chiral perturbation theory. The
results suggest
that the critical temperature $T_c$ for the chiral transition is larger
in the presence of a constant magnetic field, in agreement with 
most model calculations but in disagreement with recent lattice
calculations.

\end{abstract}

\maketitle


\section{Introduction}
Chiral perturbation theory (ChPT)
provides a systematic framework for calculating properties
of QCD at low energies~\cite{vinbjerg,gasser1,gasser12,bijn1}.
ChPT is not an expansion in powers of some small coupling constant, but it
is a systematic expansion in powers of momenta $p$ where a derivative
counts as one power and the quark masses count as two powers.
Chiral perturbation theory is a nonrenormalizable quantum field theory in the 
old sense of the word. This means that a calculation at a given order
in momentum $p$, requires that one adds higher-order operators in order
to cancel the divergences that arise in the 
calculation at that order. This implies
that one needs more and more couplings and therefore more experiments
to determine them. However, this poses no problem, as long as one is
content with finite precision. This is is the essence of effective
field theory~\cite{lepage}. 
The chiral Lagrangian that describes the (pseudo)Goldstone bosons
is uniquely determined by the global symmetries of QCD and the
assumption of symmetry breaking. The Lagrangian ${\cal L}_{\rm eff}$
consists of a string of terms that involve an increasing number
of derivatives or quark mass factors, each multiplied by a low-energy
constant (LEC) $l_i$. 
However, QCD is a confining and strongly interacting
theory at low energies. Thus the couplings $l_i$ of the chiral
Lagrangian cannot be calculated from QCD. Instead, the
couplings are fixed by experiments.

The thermodynamics of a pion gas using ChPT was studied in detail in 
a series of papers 25 years ago~\cite{gasser2,gasser3,gerber}.
The thermal pion mass and the
thermal pion decay constant were evaluated at leading order (LO), while
the pressure
and the temperature dependence of the quark condensate were calculated
to next-to-next-to-leading order (NNLO) in ChPT.
In the chiral limit, this expansion is controlled by the parameter
$T^2/8F_{\pi}^2$, where $F_{\pi}$ is the pion decay constant.
In this paper, we present calculations of 
the pion masses $M_{\pi^0}$ and $M_{\pi^{\pm}}$ as well as
the decay constants $F_{\pi^0}$ and $F_{\pi^{\pm}}$
to leading order, and the free
energy and the quark condensate to next-to-leading order (NLO) in 
ChPT
in the presence of a constant magnetic background $B$.
The details of the calculations will be presented elsewhere~\cite{jensoa}.

QCD in external magnetic fields has received a lot of attention in recent
years due to its relevance 
in several physical situations. For example, 
large magnetic fields exist inside ordinary
neutron stars as well as magnetars~\cite{neutron}. 
In the latter case, the cores
may be color superconducting and so it is important to study the effects
of external magnetic fields in this 
phase~\cite{mark,gorbar,qcdmag1,qcdmag2,fh,jorge,chinese,qcdmag3}.
Similarly, it has been suggested that strong magnetic fields are created
in heavy-ion collisions at the Relativistic Heavy-Ion Collider (RHIC) and 
the Large Hadron Collider (LHC) and that these play an important 
role~\cite{harmen1}.
In this case, the magnetic field  strength 
has been estimated to be up to $B\sim 10^{19}$ Gauss, which
corresponds to $|qB|\sim6M_{\pi}^2$, where $|q|$ is 
the electric charge of the pion.
Even larger fields could be reached due to the
effects of event-by-event fluctuations, see for example~\cite{china}.
This has spurred the interest in studying QCD in external fields.
At zero baryon chemical potential this can be done from first principles
using lattice simulations and some recent result
are found in~\cite{sanf,negro, budaleik,gunnar0}.

Chiral perturbation theory has been used to study the
quark condensate in strong magnetic fields at zero 
temperature~\cite{smilga,shusp,cptB,werbos} and finite 
temperature~\cite{agacond}.
In Ref.~\cite{agam}, the leading thermal corrections to $M_{\pi^0}$ and
$F_{\pi^0}$ in a magnetic background
were computed.
In Ref.~\cite{chiralB}.
the quark-hadron phase transition
was studied using ChPT to calculate the free energy at leading order.
The effects of external magnetic fields 
on the chiral transition have been studied in detail
using the NJL model~\cite{klev,shovkovy+,gorbie,klim,hiller,boomsma2,chat,avan,frasca,rabbi},
the Polyakov-loop extended NJL model~\cite{pnjlgat,pnjlkas},
the quark-meson model~\cite{fraga1,frasca,rabbi,rashid,anders}, the 
(P)QM model~\cite{fragapol,skokov}, 
the linear sigma model~\cite{duarte},
and the MIT bag model~\cite{fragamit}.

\section{Chiral perturbation theory}
As explained in the introduction, chiral perturbation theory is
a low-energy effective field theory that can be used to
systematically calculate physical quantities as a power series in momentum.
The effective Lagrangian is given by an infinite string
of operators involving an increasing number of derivatives or quark masses.
Schematically, we can write 
${\cal L}_{\rm eff}={\cal L}^{(2)}+{\cal L}^{(4)}+{\cal L}^{(6)}+...$
where the superscript indicates the powers of momentum. 
The leading term is given by
\bqa\nonumber
{\cal L}^{(2)}&=&{1\over4}F^2{\rm Tr}
\left[
(D_{\mu}U)^{\dagger}(D_{\mu}U)
-M^2(U+U^{\dagger})\right]
\;,
\\ &&
\label{lo}
\eqa
which is simply is the Lagrangian of the nonlinear sigma model.
Here $U=\exp[i\tau_i\pi_i/F]$ is a unitary $SU(2)$ matrix, where
$\pi_i$ are the pion fields and $\tau_i$ are the Pauli spin matrices.
The low-energy constants $M$ and $F$ are the
tree-level values for the pion mass $M_{\pi}$ and the pion decay 
constant $F_{\pi}$, respectively.
Moreover $D_{\mu}$ is the covariant derivative. 
\begin{widetext}
By expanding the
Lagrangian ${\cal L}^{(2)}$
to fourth order in the pion fields $\pi^i$, we obtain
\bqa\nonumber
{\cal L}^{(2)}&=&-F^2M^2+
{1\over2}\left(\partial_{\mu}\pi^0\right)^2
+{1\over2}M^2(\pi^0)^2
+(\partial_{\mu}+iqA_{\mu})\pi^+(\partial_{\mu}-iqA_{\mu})\pi^-
+M^2\pi^+\pi^-
\\ && \nonumber
-{M^2\over24F^2}\left[
(\pi^0)^2+2\pi^+\pi^-
\right]^2
+{1\over6F^2}\left[
-2(\pi^0)^2(\partial_{\mu}\pi^+)(\partial_{\mu}\pi^-)
-2\pi^+\pi^-(\partial_{\mu}\pi^0)^2
+[\partial_{\mu}(\pi^+\pi^-)]^2
\right. \\&& \left.
+2\pi^0[\partial_{\mu}\pi^0][\partial_{\mu}(\pi^+\pi^-)]
-4\pi^+\pi^-(\partial_{\mu}\pi^+)(\partial_{\mu}\pi^-)
\right]
\;,
\label{l2}
\eqa
where we have defined the complex pion fields
as $\pi^{\pm}={1\over\sqrt{2}}(\pi_1\pm i\pi_2)$ and $A_{\mu}=B\delta_{\mu 2}x_1$.
Similarly, expanding ${\cal L}^{(4)}$ to second order in the pion fields 
yields~\cite{shusp}
\bqa\nonumber
{\cal L}^{(4)}&=&
{1\over4}F_{\mu\nu}^2+
{2l_5\over F^2}(qF_{\mu\nu})^2\pi^+\pi^-
+{2il_6\over F^2}qF_{\mu\nu}\left[
(\partial_{\mu}\pi^-)(\partial_{\nu}\pi^+)
+iqA_{\mu}\partial_{\nu}(\pi^+\pi^-)
\right]
+(l_3+l_4){M^4\over F^2}(\pi^0)^2
\\ &&
+2(l_3+l_4){M^4\over F^2}\pi^+\pi^-
+l_4{M^2\over F^2}(\partial_{\mu}\pi^0)^2+
2l_4{M^2\over F^2}
(\partial_{\mu}+iqA_{\mu})\pi^+(\partial_{\mu}-iqA_{\mu})\pi^-
\;,
\label{l4}
\eqa
where $F_{\mu\nu}=\partial_{\mu}A_{\nu}-\partial_{\nu}A_{\mu}$ is the 
field strength tensor.
\end{widetext}
The Lagrangian ${\cal L}^{(6)}$ is very complicated as it contains
more than 50 terms for $SU(2)$~\cite{bijn1}. However, 
only one term is relevant for the
present problem~\cite{shusp,werbos}, namely
\bqa
\label{l6}
{\cal L}^{(6),\rm relevant}=-4c_{34}M^2(qF_{\mu\nu})^2\;.
\eqa
We have used the parametrization $U=e^{i\pi^i\tau_i/F}$, where $\tau_i$
are the Pauli matrices.
This parametrization is different from the one used 
in~\cite{shusp,cptB,werbos} 
and so the expressions 
for ${\cal L}$ also differ. However, we get identical results for physical
quantities independent of parametrization.
Moreover, we note that flavor symmetry is broken in an external electromagnetic
field due to the different charges of the $u$ and the $d$ quarks.
In particular, the $SU(2)_A$ symmetry is broken down to $U(1)_A^3$, which
corresponds to the rotation of the $u$ and $d$ quarks by opposite angles.
The formation of a quark condensate breaks this Abelian symmetry and gives
rise to a Goldstone boson, namely the neutral pion.
The charged pions are therefore no longer Goldstone modes. In fact, the
presence of the external electromagnetic field allows for an effective
mass term even when $M=0$, cf. the second and third term in Eq.~(\ref{l4}).

The chiral Lagrangian comes with a number of undetermined parameters or
low-energy constants $l_i$. These parameters can be determined by 
experiments; however, loop corrections involve renormalization of them.
The relation between the bare and renormalized parameters 
can be expressed as
\bqa
l_i=-{\gamma_i\over2(4\pi)^2}\left[
{1\over\epsilon}+1-\bar{l}_i
\right]\;,
\eqa
where $\gamma_i$ are coefficients 
and $\bar{l}_i$ are scale-independent parameters~\cite{gasser1}, i. e.
they are the renormalized couplings evaluated at the 
renormalization
scale $\Lambda=M$. 
In the present calculations, we
need $\gamma_3=-{1\over2}$, $\gamma_4=2$, $\gamma_5=-{1\over6}$, and 
$\gamma_6=-{1\over3}$~\cite{gasser1,gasser12}.

\section{Pion masses and pion decay constants}
The pion masses $M_{\pi^0}$ and $M_{\pi^{\pm}}$ are
defined by the position of the pole of the propagator.
At leading order, their expressions are divergent and require
renormalization of the parameters $l_3$, $l_5$, and $l_6$.
The result is
\begin{widetext}
\bqa
M_{\pi^0}^2&=&M^2_{\pi}\left[1-
{1\over(4\pi)^2F^2}\left(
I_B(M)+{1\over2}J_1(\beta M)T^2
-J_1^B(\beta M)|qB|
\right)
\right]\;,
\label{mpi0}
\\ 
M^2_{\pi^{\pm}}&=&M^2_{\pi}\left[1+{T^2\over2(4\pi)^2F^2}J_1(\beta M)\right]
+{(qB)^2\over3(4\pi)^2F^2}(\bar{l}_6-\bar{l}_5)
\;,
\label{mpip}
\eqa
\end{widetext}
where the pion mass $M_{\pi}^2$ in the vacuum is given by 
\bqa
M_{\pi}^2&=&M^2\left[1-\mbox{$M^2\over2(4\pi)^2F^2$}\bar{l}_3\right]\;,
\eqa
the function $I_B(M)$ is defined by
\bqa\nonumber
I_B(M)&=&M^2\log{M^2\over2|qB|}-M^2-2\zeta^{(1,0)}(0,\mbox{$1\over2$}+x)|qB|
\;,
\\ &&
\eqa
where $\zeta(q,s)=\sum_{m=0}^{\infty}(q+m)^{-s}$ is the Hurwitz zeta-function
and $x={M^2\over2|qB|}$. The thermal integrals are
\bqa
J_1(\beta M)&=&8\beta^{2}
\int_0^{\infty}{p^{2}dp\over\sqrt{p^2+M^2}}
{1\over e^{\beta\sqrt{p^2+M^2}}-1}\;,\\ \nonumber
J_1^B(\beta M)&=&8\sum_{m=0}^{\infty}\int_0^{\infty}
{dp\over\sqrt{p^2+M_B^2}}{1\over e^{\beta\sqrt{p^2+M^2_B}}-1}
\;,
\\ &&
\label{j1mb}
\eqa
where $M^2_B=M^2+(2m+1)|qB|$ and $m$ denotes the $m$th Landau level.

In order to calculate the pion decay constant, we need to evaluate the
matrix elements
$\langle0|{\cal A}_{\mu}^{0}|\pi^{0}\rangle$ and
$\langle0|{\cal A}_{\mu}^{\pm}|\pi^{\mp}\rangle$, where
${\cal A}_{\mu}^{0}$ and ${\cal A}_{\mu}^{\pm}$ are the axial currents for
$\pi^0$ and $\pi^{\pm}$.
At zero magnetic field, these are identical, but there are 
two pion decay constants at finite temperature;
one for the
time component and one for the spatial component of ${\cal A}_{\mu}$
since Lorentz invariance is broken.
The difference between them is an order-$p^4$ effect
i. e. appears at the two-loop level~\cite{pisarskipions} 
and this is beyond the scope of this
paper.
The matrix elements are
proportional to $iP_{\mu}$ and the prefactors are denoted by 
$F_{\pi^0}$ and $F_{\pi^{\pm}}$, respectively.
The expressions are divergent and require renormalization of
$l_4$ and the renormalized result is
\begin{widetext}
\bqa
\label{fpi00}
F_{\pi^0}&=&F_{\pi}\left[1+{1\over(4\pi)^2F^2}
\left(
I_B(M)-
J_1^B(\beta M)|qB|\right)\right]
\;, \\ 
F_{\pi^{\pm}}&=&F_{\pi}\left[1+{1\over2(4\pi)^2F^2}
\left(
I_B(M)-
J_1(\beta M)T^2-J_1^B(\beta M)|qB|\right)\right]\;,
\eqa
\end{widetext}
where the pion decay constant $F_{\pi}$ in the vacuum is 
\bqa
F_{\pi}=\left[
1+{M^2\over(4\pi)^2F^2}\bar{l}_4\right]\;.
\eqa
Note that $F_{\pi^0}$ differs from $F_{\pi^{\pm}}$
in a magnetic field.
The reason is that the loop corrections to the former involve
charged pions only, while loop corrections to the latter
involve both neutral and charged pion~\cite{jensoa}. 

\section{Free energy and quark condensate}
We are interested in the contributions to the free energy ${\cal F}$
that are due to a nonzero magnetic field and finite temperature.
We therefore write the contribution to the free energy 
at the $n$th loop order, ${\cal F}_n$,  as
a sum of three terms: 
${\cal F}_n={\cal F}^{\rm vac}_n+{\cal F}^{B}_n+{\cal F}^{T}_n$,
where ${\cal F}^{\rm vac}_n$ is the free energy in the vacuum, i. e. 
$B=T=0$, ${\cal F}^{B}_n$ is the zero-temperature contribution due
to a finite magnetic field, and ${\cal F}^{T}_n$ is the finite-temperature
contribution. The strategy is to isolate the term
${\cal F}^{\rm vac}_n$ and subtract it from ${\cal F}_n$.
This term contains ultraviolet divergences which are removed by 
renormalization of the low-energy constants of the chiral Lagrangian
and the renormalized ${\cal F}^{\rm vac}_n$ represents
the vacuum energy of the theory.
The term ${\cal F}^{B}_n$ generally contains
ultraviolet divergences as well and it is rendered finite
by renormalizing the $l_i$s. In the present case, 
$\bar{l}_5$ and $\bar{l}_6$ in Eq.~(\ref{l4}), and
$c_{34}$ in Eq.~(\ref{l6}) 
require renormalization.
If we express the contributions ${\cal F}_1^B$ and ${\cal F}_1^T$ 
in terms of 
the physical pion masses $M_{\pi^0}(0)$, Eq.~(\ref{mpi0}), 
and $M_{\pi^{\pm}}(0)$, Eq.~({\ref{mpip}}),
at zero temperature,
instead of $M$, most of the dependence on the constants $\bar{l}_i$s
cancels in the expressions for ${\cal F}_{1+2}^B$ and 
${\cal F}_{1+2}^T$. After a lenghty calculation, one finds~\cite{jensoa}
\begin{widetext}
\bqa
{\cal F}^{B}_{1+2}&=&
{M_{\pi^{\pm}}^4(0)\over2(4\pi)^2}
\left[1-2\log{M_{\pi^{\pm}}^2(0)\over2|qB|}\right]
+{4(qB)^2\over(4\pi)^2}\zeta^{(1,0)}(-1,\mbox{$1\over2$}+x_{\pi^{\pm}})
+{(qB)^2\over6(4\pi)^2}\log{\Lambda^2\over2qB}
-{(qB)^2\over(4\pi)^4F^2}
\bar{d}(M^2)M^2
\;,
\\ \nonumber
{\cal F}^{T}_{1+2}&=&
-{1\over2(4\pi)^2}\left[J_0(\beta M_{\pi^0}(0))T^4
+2J_0^B(\beta M_{\pi^{\pm}}(0))|qB|T^2\right]
+{M^2\over8(4\pi)^4F^2}\bigg[
-J_1^2(\beta M)T^4
\\ &&
+4J_1(\beta M)J_1^B(\beta M)T^2|qB|
\bigg\}
\;,
\eqa
\end{widetext}
where 
\bqa
\bar{d}(M^2)
=8(4\pi)^4c_{34}^r-{1\over3}(\bar{l}_6-\bar{l}_5)\log{M^2\over\Lambda^2}\;,
\eqa
$x_{\pi^{\pm}}={M_{\pi^{\pm}}^2(0)\over2|qB|}$,
and $\Lambda$ is the renormalization scale.
The term ${(qB)^2\over6(4\pi)^2}\log{\Lambda^2\over2qB}$
arises from wave function renormalization of the term ${1\over{2}}B^2$
in the tree-level 
expression for the free energy ${\cal F}_0={1\over{2}}B^2-F^2M^2$.
It cancels a logarithmic divergence in ${\cal F}_1^B$
proportional to $(qB)^2$. This term is typically ignored in the 
literature since it is independent of $T$ and the parameters of 
the chiral Lagrangian.

We note that the NLO correction to the free energy
in the chiral limit ($M=0$) does not vanish 
since $\pi^{\pm}$ are no longer Goldstone modes and
$M_{\pi^{\pm}}(0)$ is nonzero. This is 
in contrast to the case of zero magnetic field~\cite{gasser2,gasser3,gerber}.

At finite temperature, the quark condensate is
\bqa
\langle\bar{q}q\rangle=
\langle0|\bar{q}q|0\rangle\left(
1-{c\over F^2}{\partial({\cal F}^T+{\cal F}^B)\over\partial M_{\pi}^2}\right)
\;,
\eqa
where 
the constant $c$ is defined by~\cite{gerber}
\bqa
c=-F^2{\partial M_{\pi}^2\over\partial m_q}\langle0|\bar{q}q|0\rangle^{-1}\;.
\eqa
Here $m_q$ is the quark mass.
In the chiral limit, we have $c=1$. In that case, the
quark condensate reduces to
\begin{widetext}
\bqa\nonumber
\langle\bar{q}q\rangle
&=&
\langle0|\bar{q}q|0\rangle
\left\{
1+{|qB|\over(4\pi)^2F^2}I_B(M_{\pi^{\pm}}(0))
+{(qB)^2\over(4\pi)^4F^4}
\bar{d}(|qB|)
-{1\over2(4\pi)^2F^2}\Big(J_1(0)T^2+2J_1^B(\beta M_{\pi^{\pm}}(0))|qB|\Big)
\right. \\ &&\left.
+{T^2\over8(4\pi)^4F^4}\Big(
J_1^2(0)T^2-4J_1(0)J_1^B(0)|qB|
+4\log2\,J_1(0)|qB|\Big)
\right\}\;.
\label{quarkie}
\eqa
\end{widetext}
This is the main result of the present paper and will be discussed in the
next section.

\section{Results and discussion}
We first notice that we in the limit $B\rightarrow0$ recover 
the temperature dependence of
$M_{\pi}$, $F_{\pi}$, ${\cal F}$, and $\langle\bar{q}q\rangle$ 
as in ~\cite{gasser2,gasser3,gerber}. 
Similarly, we obtain the $T=0$ result for the
free energy and the $B$ dependence of the quark condensate as 
in~\cite{shusp,cptB,werbos}. The results (\ref{mpi0}) for $M_{\pi^0}^2$
and (\ref{fpi00}) for $F_{\pi^0}$ were first obtained in~\cite{agam}.
The neutral pion decay constant depends on the magnetic field, which
perhaps is unexpected. However, it is simply due to
a cubic term $(\pi^+\pi^-)\partial_{\mu}\pi^0$ in the expression for
the axial current ${\cal A}_{\mu}^0$ and gives rise to a 
charged pion loop~\cite{agam,jensoa}.

We also notice that the 
temperature dependence of the charged pion mass is the same
as for vanishing magnetic field. 
The only difference is a temperature-independent
constant proportional to $(qB)^2/F^2$ arising from the second and third
terms in Eq.~(\ref{l4}). Thus the charged pions are massive excitations
even in the limit when the quark mass $m_q$ goes to zero.
This simply reflects that only the neutral pion is a Goldstone mode in 
an external electromagnetic field.

The temperature dependence of $M^2_{\pi^{\pm}}$
may seem surprising at first since
there are loop corrections to $M^2_{\pi^{\pm}}$ involving charged
pion loops. 
However, these loop corrections cancel after having
taken appropriately into account wave function renormalization of
the charged pion fields~\cite{jensoa}.

In the remainder we focus on the chiral limit.
In this case there are two dimensionless ratios, namely $|qB|/T^2$ and
$T^2/F^2$. 
The integrals $J_n^B$ are functions only of the 
dimensionless ratio $|qB|/T^2$. 
It is straightforward to show that $J_1T^2\geq J_1^B|qB|$ for all values of $B$
and $T$. This implies that the pion decay constants $F_{\pi^0}$
and $F_{\pi^{\pm}}$ are larger than $F_{\pi}$.
Moreover, for small values of $|qB|$, i.e. for $|qB|\ll T^2$, 
we can calculate the first
corrections due to nonzero $B$ as a power series in $\sqrt{|qB|}/T$.
One finds 
\begin{widetext}
\bqa
F_{\pi^0}&=&F_{\pi}\left(1+{|qB|\log2\over(4\pi)^2F^2}
-{T^2\over12F^2}+{5\sqrt{|qB|}T\over48\pi F^2}+...\right)\;,\\
F_{\pi^{\pm}}&=&F_{\pi}\left(1
+{|qB|\log2\over2(4\pi)^2F^2}
-{T^2\over12F^2}+{5\sqrt{|qB|}T\over96\pi F^2}+...\right)\;.
\eqa
Similarly, we can expand the quark condensate around 
$|qB|=0$ and 
obtain the first correction proportional to $\sqrt{|qB|}/T$:
\bqa
\langle\bar{q}q\rangle&=&
\langle0|\bar{q}q|0\rangle\left(1+{|qB|\log2\over(4\pi)^2F^2}-{T^2\over8F^2}
+{5\sqrt{|qB|}T\over48\pi F^2}+...\right)\;. 
\eqa
In the limit $|qB|\rightarrow\infty$, $J_1^B\rightarrow0$ 
since the terms in the sum in Eq.~(\ref{j1mb}) 
are effectively Boltzmann suppressed. 
Eq.~(\ref{mpi0}) then shows that the dominant contribution to $M_{\pi^0}^2$
goes like $-I_B(0)=-|qB|\log2$ and so 
$M_{\pi^0}^2$ eventually turns negative which obviously
is unphysical. 
From Eq.~(\ref{fpi00}), we see
that $F_{\pi^0}$ becomes temperature independent.

\begin{figure}[htb]
\includegraphics[width=8cm]{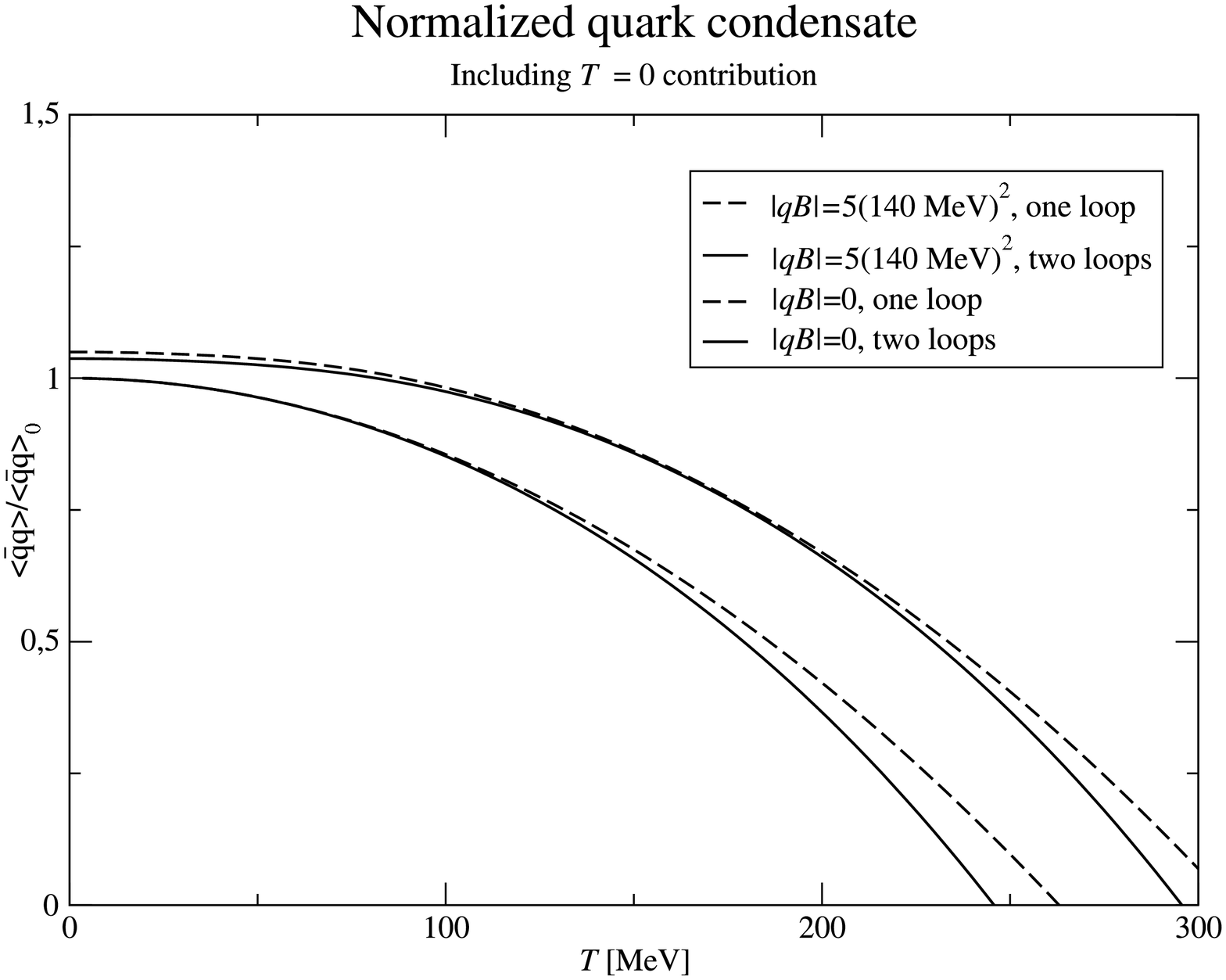}
\includegraphics[width=8cm]{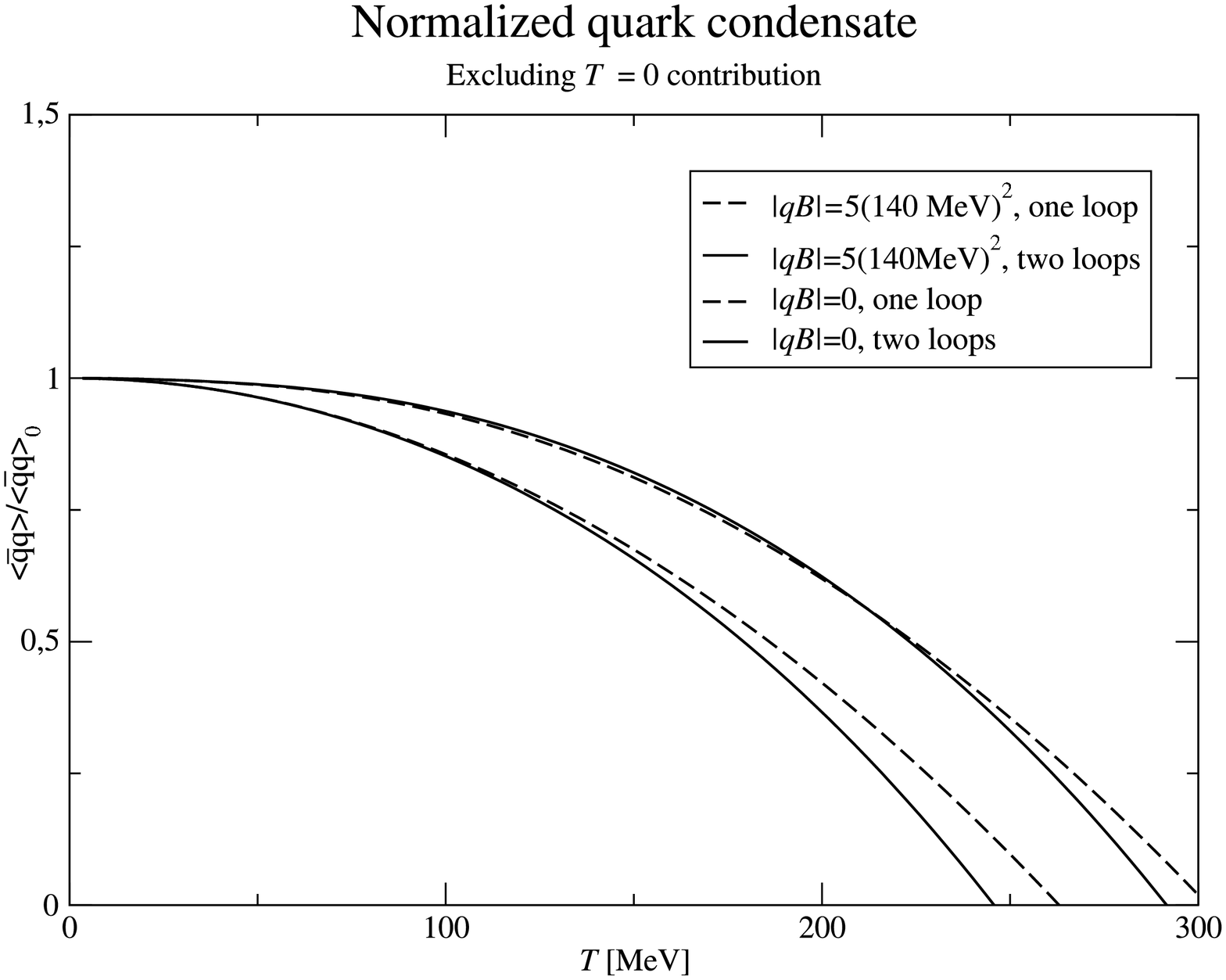}
\caption{
Temperature dependence of the quark condensate 
including the $T=0$ contribution
normalized
to its vacuum value 
$qB=5\,\,(140\,\,{\rm MeV})^2$ at LO and NLO in chiral perturbation
theory.
For comparison, we show the LO and NLO results for $qB=0$ as well.}
\label{condensate}
\end{figure}

\end{widetext}

In Fig.~\ref{condensate} (left panel), 
we show the quark condensate Eq.~(\ref{quarkie})
as a function of temperature for $|qB|=5\,(\,140 {\rm\,\,MeV})^2$ 
at LO and NLO in chiral perturbation theory 
including the $T=0$ contribution.
For comparison, we also show the quark condensate for $|qB|=0$.
We are using the experimental value $F_{\pi}=93$ MeV and
$\bar{l}_6-\bar{l}_5=3.0\pm0.3$~\cite{exp1,exp2}.
There is a large uncertainty in the constant $\bar{d}(|qB|)$ and its
value is consistent with zero and we choose
this value for simplicity.
In Fig.~\ref{condensate} (right panel), we show the quark 
condensate Eq.~(\ref{quarkie})
except that we have excluded the zero-temperature contribution.
We do this to disentangle the effects of the magnetic field at $T=0$
and the finite-temperature effects.
We notice that the LO and NLO results for the 
condensate in both cases are very close
to each other in the entire temperature range. 
In fact, the LO and NLO curves lie significantly closer than do the
corresponding curves for $B=0$. This suggests that chiral perturbation
theory converges at least as well in the presence of a magnetic field.

The quark condensate for vanishing $B$ goes faster to zero than 
it does in the presence of a magnetic field. 
This effects is caused by two separate mechanisms.
Firstly, there is the enhanced quark condensate at $T=0$, which
to leading order is determined by the function $I_B(M)$.
This is the well-known enhancement of the chiral condensate
in the presence of a magnetic field.
Secondly, there are finite-temperature corrections.
The basic effect here is that $J_1^B$ is a decreasing function of $B$ and
thus $J_1T^2> J_1^B|qB|$ for all $B>0$.
Using this inequality, it is straightforward to show that the  
decrease of the quark condensate~(\ref{quarkie}) due to thermal
effects is smaller for nonzero $B$.
The 
two separate effects are clearly demonstrated if one compares the two panels in 
Fig.~\ref{condensate}.

Comparing the results for the condensate for $B=0$ and 
$|qB|=5\,(140\,{\rm MeV})^2$, it is clear
the effects of the magnetic field are quantitatively
large. 
This is due to a very strong magnetic field.
For smaller values of $|qB|$, the gaps between the two sets of curves
will be smaller too.
The calculations indicate that the
critical temperature $T_c$ for the chiral transition is higher
in a nonzero magnetic field. Of course, this conclusion is cautious since
the behavior of the quark condensate in the vicinity of $T_c$
is beyond the reach of chiral perturbation theory.
This result is in line with most model 
calculations, both mean-field type~\cite{fragapol,pnjlgat,pnjlkas,duarte} 
and beyond~\cite{anders,skokov}. 
Model calculations that seem to indicate a decrease of $T_c$
as a function of magnetic field
can be found in 
Refs.~\cite{chiralB,fragamit}.

In Ref.~\cite{chiralB}, the authors use ChPT at leading order
to investigate the quark-hadron 
phase transition as a function of the magnetic field at the physical point.
They compare the pressure in the hadronic phase with that of the quark-gluon
plasma phase for an ideal gas of quarks and gluons, and subtracting
the vacuum energy due to a nonzero
gluon condensate. 
For weak magnetic fields, the transition is first order. The line
of first-order transitions ends at a critical point. From this temperature
onwards, the transition is a crossover.
The critical temperature defined this way is a decreasing function of
$B$.
Typically, however, the critical temperature is determined by the 
behavior of the quark condensate. At the physical point, the condensate never
vanishes and the transition is a crossover.
The crossover temperature is often defined by the inflection point
of $\langle\bar{q}q\rangle$ as a function of temperature.

D'Elia {\it et al} have carried out lattice simulations in a constant
magnetic background at zero chemical potential~\cite{sanf,negro}.
They explored various constituent quark masses corresponding to a pion mass
of $200-480$ MeV and different magnetic fields, up to $|qB|\sim20$ $M_{\pi}^2$
for the lightest quark masses.
For these values of the pion mass, 
they found that
there is a slight increase in the critical temperature $T_c$ for the
chiral transition. These results have been confirmed by
Bali {\it et al}~\cite{budaleik,gunnar0,gunnar}.
The same group has alo carried out 
lattice simulations for physical values of the pion mass, 
i.e. $M_{\pi}=140$ MeV. Their results which are extrapolated to the continuum
limit show that the 
critical temperature is a decreasing function of the magnetic field
\cite{budaleik,gunnar0,gunnar}. Hence the critical temperature 
for fixed $|qB|$ as a function
of the quark mass is nontrivial. 
This is in stark contrast to most model calculations
that imply an increasing critical temperature as a functions of $B$.
This is irrespective of whether one goes beyond mean field or not.
The discrepancy is perhaps somewhat surprising since 
at $T=0$, the lattice results confirm the magnetic catalysis predicted by
model calculations.

In conclusion, we have used chiral perturbation theory to calculate
the pion masses, the decay constants, the free energy and the quark condensate
at finite temperature in a magnetic background.
Given the conflicting results for $T_c$ as a function of $B$
of various model calculations and lattice 
calculations, clearly more work needs to be done.
\section*{Acknowledgments}
The author would like to thank G. Bali,
G. Endr\H{o}di, and F. Bruckmann
for useful discussions on their lattice simulations,
and N. O. Agasian for discussions on ChPT.
He thanks the Niels Bohr International Academy and the Discovery Center for 
kind hospitality during the course of this work.

\end{document}